\documentclass[aip,reprint,citeautoscript,groupedaddress,noeprint]{revtex4-2}
\usepackage{amsmath}
\usepackage{bm}
\usepackage{graphicx,color}
\usepackage{stix}
\usepackage{subfigure}
\usepackage{physics}
\usepackage{array}
\usepackage[usenames,dvipsnames]{xcolor}

\newcommand{\eri}[2]{{\left( #1 \middle| #2 \right)}}

\newcommand*{\veck}{{\mathbf{k}}}

\newcommand*{\vecr}{{\mathbf{r}}}

\newcommand*{\vecq}{{\mathbf{q}}}
\newcommand*{\vecI}{{\mathbf{I}}}
\newcommand*{\vecPi}{{\mathbf{\Pi}}}
\newcommand*{\vecSig}{{\mathbf{\Sigma}}}

\newcommand*{\occ}{{\mathrm{occ}}}
\newcommand*{\vir}{{\mathrm{vir}}}

\newcommand*{\gw}{{\textit{G}\textsubscript{0}\textit{W}\textsubscript{0}}}
\newcommand*{\GW}{\textit{GW}}

\definecolor{myblue}{rgb}{0,0,1}
\usepackage[breaklinks=true,colorlinks=true,linkcolor=myblue,urlcolor=myblue,citecolor=myblue]{hyperref}

\begin{document}
\title{Gaussian-Based Quasiparticle Self-Consistent \textit{GW} for Periodic Systems}
\author{Jincheng Lei}
\author{Tianyu Zhu}
\email{tianyu.zhu@yale.edu}
\affiliation{Department of Chemistry, Yale University, New Haven, CT 06520}

\begin{abstract}
We present a quasiparticle self-consistent \textit{GW} (QSGW) implementation for periodic systems based on crystalline Gaussian basis sets. Our QSGW approach is based on a full-frequency analytic continuation \textit{GW} scheme with Brillouin zone sampling and employs the Gaussian density fitting technique. We benchmark our QSGW implementation on a set of weakly-correlated semiconductors and insulators as well as strongly correlated transition metal oxides including MnO, FeO, CoO, and NiO. Band gap, band structure, and density of states are evaluated using finite size corrected QSGW. We find that although QSGW systematically overestimates band gaps of tested semiconductors and transition metal oxides, it completely removes the dependence on the choice of density functionals and provides more consistent prediction of spectral properties than \gw~across a wide range of solids. This work paves the way for utilizing QSGW in \textit{ab initio} quantum embedding for solids.
\end{abstract}

\maketitle

\section{Introduction}
The many-body perturbation theory, or \GW, is one of the most promising electronic structure methods for simulating materials spectra. An approximation to solve Hedin's equations~\cite{Hedin1965}, the \GW~theory~\cite{Hanke1979,Strinati1982,Hybertsen1986,Aryasetiawan1998,Golze2019a} is formally defined as a self-consistent Green's function approach that computes the self-energy $\Sigma$ from multiplication of one-particle Green's function $G$ and screened Coulomb interaction $W$: $\Sigma = i G W$, where $W$ is computed at the level of random phase approximation (RPA)~\cite{Ren2012a}. In practice, the most widely-used \GW~form is the one-shot \gw~approximation with no self-consistency, due to its lower computational cost. Starting from density functional theory (DFT)~\cite{Kohn1965} with local density approximation (LDA) or generalized gradient approximation (GGA), \gw~has been shown to yield accurate band gaps for weakly-correlated semiconductors~\cite{Shishkin2006,Gonze2009,Deslippe2012,Marini2009,Govoni2015,gulans2014,Huser2013,Friedrich2010,Zhu2021a,Ren2021}. However, the accuracy of the \gw~method has a dependence on the starting choice of density functionals and \gw@LDA/GGA may be qualitatively wrong for strongly correlated materials~\cite{Bruneval2013,Korzdorfer2012,Chen2014a}.

Several levels of self-consistency have been proposed to alleviate the starting-point dependence. The fully self-consistent \GW~(scGW) approach iteratively solves Dyson's equation $G=G_0+G_0\Sigma G$ by updating Green's function ($G$), screened Coulomb interaction ($W$), and self-energy ($\Sigma$) until convergence, but it is very computationally demanding~\cite{Caruso2012, Caruso2013, Kutepov2017a,Grumet2018,Yeh2022}. Thus, simplified \GW~schemes such as eigenvalue self-consistent \GW~(evGW) and quasiparticle self-consistent \GW~(QSGW) have been developed. In evGW, one self-consistently updates the orbital energies that enter into $G$ and $W$, but does not update the orbitals~\cite{Hybertsen1986,Shishkin2007}. On the other hand, QSGW is a formalism that determines the optimum effective Hamiltonian $V^\mathrm{eff}$ for the quasiparticle (QP) equation by minimizing the \GW~perturbative correction to $V^\mathrm{eff}$~\cite{Faleev2004a,VanSchilfgaarde2006b,Kotani2007b,Shishkin2007b,Bruneval2006,Bruneval2014}. In QSGW, both quasiparticle energies and orbitals are optimized in a self-consistent manner, although the frequency dependence in self-energy is simplified to a static effective Hamiltonian. Originally introduced by Faleev, Van Schilfgaarde, and Kotani~\cite{Faleev2004a}, QSGW was shown to yield reliable results for many semiconductors and transition metal oxides. In addition, partially self-consistent \GW~are also used to accelerate evGW and QSGW calculations, such as evGW0 and QSGW0 (i.e., $W$ is kept at the DFT level during self-consistency). Among all these formulations, scGW and QSGW are the only \GW~methods that completely removes the starting point dependence on DFT functionals.

For periodic sytems, QSGW has been predominantly implemented using plane-wave (PW) basis. To represent the sharp core electron density, either pseudopentials are utilized~\cite{Bruneval2006,Chen2015} or the PW basis is augmented by other functions, such as linear muffin-tin orbitals (LMTO)~\cite{Kotani2007b}, projector augmented wave (PAW)~\cite{Shishkin2007b}, and linear augmented
plane waves (LAPW)~\cite{Deguchi2016a}. Although highly efficient QSGW implementations exist, a drawback for employing PW basis is the need for a large number of virtual bands to converge the screened interaction. 

In recent years, molecular QSGW implementations adopting Gaussian atomic orbitals have appeared~\cite{Kaplan2016a,Forster2021}. Due to the compactness of Gaussian basis, molecular QSGW calculations are normally performed without truncation of virtual space. Moreover, because local orbitals are well suited to describe rapid oscillations of core electron density, Gaussian-based QSGW has been applied to study core level binding energies of small molecules~\cite{VanSetten2018}.

Despite all these promising developments, Gaussian-based QSGW for periodic systems is not yet available. In Ref.~\cite{Zhu2021a}, one of us (T.Z.) implemented all-electron Gaussian-based \gw~approach for solids within the PySCF quantum chemistry software package~\cite{Sun2020b}. In this work, we extend the capability of this framework and develop QSGW method with Brillouin zone sampling using crystalline Gaussian basis sets. We benchmark our QSGW implementation on the band gap, band structure, and density of states of weakly-correlated semiconductors as well as strongly-correlated transition metal oxides including MnO, FeO, CoO, and NiO, and compare its performance to one-shot \gw~under the same numerical setting. 

\section{Theory and Implementation}
\subsection{Gaussian-based periodic \gw}
The Gaussian-based periodic \gw~approach was described in details in Ref.~\cite{Zhu2021a}. Here, we only summarize its main formalism based on the analytic continuation (AC) scheme, which is necessary for discussing the QSGW implementation. 

In the \gw-AC scheme~\cite{Ren2012,Wilhelm2016,Zhu2021a}, the self-energy is first computed along the imaginary frequency axis: 
\begin{equation}\label{eq:sigma1}
\Sigma(\vecr, \vecr', i\omega) = -\frac{1}{2\pi} \int_{-\infty}^{\infty} d\omega' G_0(\vecr, \vecr', i\omega+i\omega') W_0(\vecr, \vecr', i\omega'),
\end{equation}
and then analytically continued to the real axis to compute quasiparticle energies. The non-interacting DFT Green's function $G_0$ on the imaginary axis is expressed as
\begin{equation}\label{eq:G0AC}
G_0(\vecr, \vecr', i\omega) = \sum_{m\veck_m} \frac{\psi_{m\veck_m}(\vecr) \psi_{m\veck_m}^*(\vecr')}{i\omega + \epsilon_F - \epsilon_{m\veck_m}},
\end{equation}
where $\psi_{m\veck_m}(\vecr)$ and $\epsilon_{m \veck_m}$ refer to DFT molecular orbitals (MO) and orbital energies. The Fermi energy $\epsilon_F$ is taken to be the midpoint between the DFT valence band maximum and conduction band minimum energies for gapped systems. 

The screened Coulomb interaction $W_0$ is defined as
\begin{equation}\label{eq:W0}
W_0(\vecr, \vecr', i\omega) = \int d\vecr'' \varepsilon^{-1}(\vecr, \vecr'', i\omega) v(\vecr'', \vecr'),
\end{equation}
where $v(\vecr, \vecr') = |\vecr - \vecr'|^{-1}$ is the Coulomb operator, and $\varepsilon(\vecr, \vecr', i\omega)$ is the dielectric function:
\begin{equation}\label{eq:epsr}
\varepsilon(\vecr, \vecr', i\omega) = \delta(\vecr, \vecr') - \int d\vecr'' v(\vecr, \vecr'') \chi_0(\vecr'',\vecr',i\omega) .
\end{equation}
$\chi_0(\vecr,\vecr',i\omega)$ is the polarizability calculated at the level of random phase approximation (RPA):
\begin{equation}
\begin{split}
\chi_0(\vecr,\vecr',i\omega) = \frac{1}{N_\veck} & \sum_{i\veck_i}^{\occ} \sum_{a\veck_a}^\vir \bigg( \frac{\psi_{i\veck_i}^*(\vecr) \psi_{a\veck_a}(\vecr) \psi_{a\veck_a}^*(\vecr') \psi_{i\veck_i}(\vecr')}{i\omega - (\epsilon_{a\veck_a} - \epsilon_{i\veck_i}) + i\eta} \\ 
& - \frac{\psi_{i\veck_i}(\vecr) \psi_{a\veck_a}^*(\vecr) \psi_{a\veck_a}(\vecr') \psi_{i\veck_i}^*(\vecr')}{i\omega + (\epsilon_{a\veck_a} - \epsilon_{i\veck_i}) - i\eta} \bigg),
\end{split}
\end{equation}
with $i$ and $a$ labeling occupied and virtual molecular orbitals respectively.

In Gaussian-based \gw, we employ the Gaussian density fitting (GDF) technique~\cite{Sun2017b} to decompose 4-center 2-electron electron repulsion integrals (ERIs) into 3-center 2-electron tensors:
\begin{equation}
\eri{p\veck_p q\veck_q}{r\veck_r s\veck_s} = \sum_R v_{R \veck_{pq}}^{p\veck_p, q\veck_q} \cdot v_{R\veck_{rs}}^{r\veck_r, s\veck_s} .
\end{equation}
Here, $p, q, r, s$ are crystalline Gaussian atomic orbitals and $R$ denotes auxiliary periodic Gaussian functions. $\veck_{pq}$, $\veck_p$ and $\veck_q$ satisfy momentum conservation: $\veck_{pq} = \veck_p - \veck_q + N\mathbf{b}$, where $\mathbf{b}$ is a reciprocal lattice vector and $N$ is an integer. To maintain consistency with plane-wave expressions in the \GW~literature, we relabel the $v_{R \veck_{rs}}^{r\veck_r, s\veck_s}$ integral as $v_{R\vecq}^{r\veck, s\veck-\vecq}$, where $\veck$ and $\vecq$ are crystal momentum vectors.

Using GDF integrals, the matrix elements of $W_0$ are computed as
\begin{equation}\label{eq:Wmat}
\begin{split}
& \left( n\veck, m\veck-\vecq \middle| W_0 \middle| m\veck-\vecq, n'\veck \right) \\
&= \sum_{PQ} v_P^{nm} [ \vecI-\vecPi(\vecq, i\omega) ]_{PQ}^{-1} v_Q^{mn'},
\end{split}
\end{equation}
where $m, n, n'$ refer to molecular orbitals and $P, Q$ denote auxiliary orbitals. The response function $\vecPi(\vecq, i\omega)$ is computed as
\begin{equation}\label{eq:Pi}
\begin{split}
\vecPi_{PQ}(\vecq, i\omega) = \frac{2}{N_\veck} \sum_{\veck} \sum_{i}^{\occ}\sum_{a}^{\vir} v_{P}^{ia} \frac{\epsilon_{i\veck} - \epsilon_{a\veck+\vecq}}{\omega^2 + (\epsilon_{i\veck} - \epsilon_{a\veck+\vecq})^2} v_{Q}^{ai} .
\end{split}
\end{equation}

Inserting Eq.~\ref{eq:G0AC} and Eq.~\ref{eq:Wmat} into Eq.~\ref{eq:sigma1}, we obtain the self-energy matrix elements:
\begin{equation}\label{eq:sigmaAC2}
\begin{split}
\vecSig_{nn'}(\veck, i\omega) = & -\frac{1}{2\pi N_\veck} \sum_{m\vecq} \int_{-\infty}^{\infty} d\omega' \frac{1}{i(\omega+\omega')+\epsilon_F-\epsilon_{m\veck-\vecq}}  \\
&  \times \sum_{PQ} v_P^{nm} [\vecI-\vecPi(\vecq, i\omega')]_{PQ}^{-1} v_Q^{mn'} .
\end{split}
\end{equation}
The \gw~self-energy can be further partitioned into an exchange part, which is simply Hartree-Fock exchange matrix
\begin{equation}\label{eq:sigmax}
\vecSig_{nn'}^x(\veck) = -\frac{1}{N_\veck} \sum_{P\vecq} \sum_{i}^{\occ} v_{P\vecq}^{n\veck, i\veck-\vecq} \cdot v_{P(-\vecq)}^{i\veck-\vecq, n'\veck} ,
\end{equation}
plus a \gw~correlation part
\begin{equation}\label{eq:sigmaAC3}
\begin{split}
\vecSig^c_{nn'}(\veck, i\omega) = & -\frac{1}{\pi N_\veck} \sum_{m\vecq} \int_{0}^{\infty} d\omega' \frac{i\omega+\epsilon_F-\epsilon_{m\veck-\vecq}}{(i\omega+\epsilon_F-\epsilon_{m\veck-\vecq})^2 + \omega'^2}  \\
&  \times \sum_{PQ} v_P^{nm} \big[ [\vecI-\vecPi(\vecq, i\omega')]^{-1}_{PQ}-\delta_{PQ} \big] v_Q^{mn'} .
\end{split}
\end{equation}
The integration in Eq.~\ref{eq:sigmaAC3} is done using numerical grids such as modified Gauss-Legendre quadrature (with 100 grid points). After the self-energy is obtained on the imaginary axis, one analytically continues the self-energy to the real axis using Pad\'{e} approximants~\cite{Vidberg1977}. Finally, one solves the \gw~QP equation to obtain \gw~QP energies:
\begin{equation}\label{eq:qp}
\epsilon_{n\veck}^{\GW} = \epsilon_{n\veck}^{\mathrm{DFT}} + \left( \psi_{n\veck} \middle| \mathrm{Re} \ \Sigma(\epsilon_{n\veck}^{\GW}) - v^{xc} \middle| \psi_{n\veck} \right) .
\end{equation}

\subsection{Quasiparticle self-consistent \GW}
We follow Kotani \textit{et al.}~\cite{Kotani2007b} to formulate our Gaussian-based QSGW approach. The QP equation for a given system is written as
\begin{equation}
    \big[-\frac{1}{2} \nabla^2 + V_\mathrm{ext} + V_H + \mathrm{Re} [\Sigma(E_i)] \big] \Phi_i = E_i \Phi_i ,
\end{equation}
where $V_\mathrm{ext}$ and $V_H$ are nuclear potential and Hartree potential, and $E_i$ and $\Phi_i$ are QP energies and orbitals. The idea of QSGW is to map the dynamic frequency-dependent \GW~self-energy $\Sigma(\omega)$ into a static effective potential $V_\mathrm{QSGW}$, and self-consistently solve the QSGW equation
\begin{equation}
    \big[-\frac{1}{2} \nabla^2 + V_\mathrm{ext} + V_H + V_\mathrm{QSGW} \big] \psi_i = \epsilon_i \psi_i  ,
\end{equation}
where the QSGW effective potential $V_\mathrm{QSGW}$ is determined by the QSGW QP energies $\epsilon_i$ and orbitals $\psi_i$.

In literature, several ways of the mapping $\Sigma(\omega) \rightarrow V_\mathrm{QSGW}$ were proposed. The most commonly used form is
\begin{equation}
    V_\mathrm{QSGW} = \frac{1}{2} \sum_{ij} \ket{\psi_i} \big\{ \mathrm{Re} [\Sigma(\epsilon_i)] + \mathrm{Re} [\Sigma(\epsilon_j)] \big\} \langle \psi_j |,
\label{eq:modeA}
\end{equation}
referred to as ``mode A'' in Ref.~\cite{Kotani2007b}. In the meantime, a closely related form of mapping, known as ``mode B'', is defined as
\begin{equation}
    V_\mathrm{QSGW} = \sum_{i} \ket{\psi_i} \mathrm{Re} [\Sigma(\epsilon_i)] \bra{\psi_i} + \sum_{i \neq j} \ket{\psi_i} \mathrm{Re} [\Sigma(\epsilon_F)] \langle \psi_j |
\label{eq:modeB}
\end{equation}
As can be seen, the difference between ``mode A'' and ``mode B'' is in the treatment of off-diagonal elements of $V_\mathrm{QSGW}$. While the $ij$ element of $V_{\mathrm{QSGW}}$ is obtained by averaging self-energies computed at $\epsilon_i$ and $\epsilon_j$ in ``mode A'', it is taken as the self-energy computed at the Fermi energy $\epsilon_F$ in ``mode B''. Similar QP energies were obtained using these two modes in Ref.~\cite{Kotani2007b}.

When implemented using the AC scheme in Gaussian basis, we find that ``mode A'' (Eq.~\ref{eq:modeA}) leads to numerical instability and QSGW convergence issue. This is because the commonly used AC approach, such as Pad\'{e}, is incapable of producing stable off-diagonal elements of the QSGW self-energy on the real axis, especially when computed at high-lying virtual orbital energy or core orbital energy. Such behavior was also observed in a recent work by F\"orster \textit{et al}~\cite{Forster2021}. On the other hand, when implemented using ``mode B'' (Eq.~\ref{eq:modeB}), our QSGW approach becomes numerically stable since it is well known that AC performs best close to the Fermi energy, leading to improved accuracy on the off-diagonal self-energy elements. Thus, in the remaining part of this paper, all results are generated using the ``mode B'' implementation of QSGW. 

Furthermore, we implemented the direct inversion in the iterative subspace (DIIS) technique~\cite{Pulay1980} to accelerate the QSGW self-consistent iterations. Specifically, we consider a QSGW calculation to be converged when the following convergence criterion is satisfied twice consecutively:
\begin{equation}
    \frac{1}{N_\veck N}  |\gamma_i^\mathrm{QSGW} - \gamma_{i-1}^\mathrm{QSGW}| < \delta,
    \label{eq:converge}
\end{equation}
where $\gamma_i^\mathrm{QSGW}$ is the QSGW density matrix at $i$th cycle, $N$ and $N_\veck$ are the number of orbitals per unit cell and the number of $\veck$-points. The convergence threshold $\delta$ is set to be $10^{-5}$ a.u..

\section{Results}
\subsection{Computational details}
We first benchmark our Gaussian-based QSGW method on a set of semiconductors and insulators, including Si, C (diamond), SiC, BN, AlN, AlP, MgO, ZnO, as well as rare gas solids, including Ar and Ne. AlN and ZnO are in the wurtzite crystal structure, while all other solids are in the cubic structure. We employed all-electron cc-pVTZ basis set~\cite{Dunning1989,woon1993,Prascher2011} and the cc-pVTZ-RI auxiliary basis set~\cite{Hattig2005,Weigend2002a} for all elements except Ar, Ne, and Zn. For Ar and Ne, we used aug-cc-pVTZ/aug-cc-pVTZ-RI basis sets.~\cite{Kendall1992} For Zn, we used cc-pVTZ-pp/cc-pwCVTZ-PP-RI basis sets~\cite{Peterson2005,Figgen2005} to remove the Zn core electrons in the calculations. 

We also apply our QSGW approach to study transition metal oxides (TMOs) including MnO, FeO, CoO, and NiO. We simulated all TMOs in the antiferromagnetic phase and the rock salt crystal structure. Due to the larger unit cell size (two TMO units) in these materials, we adopted a smaller GTH-DZVP-MOLOPT-SR basis set in combination with GTH-HF pseudopotential.~\cite{Hartwigsen1998,Vandevondele2005} All calculations were performed with the PySCF quantum chemistry software package~\cite{Sun2020b}.

\subsection{QSGW convergence}
\begin{figure}[hbt]
\centering
\includegraphics[width=0.45\textwidth]{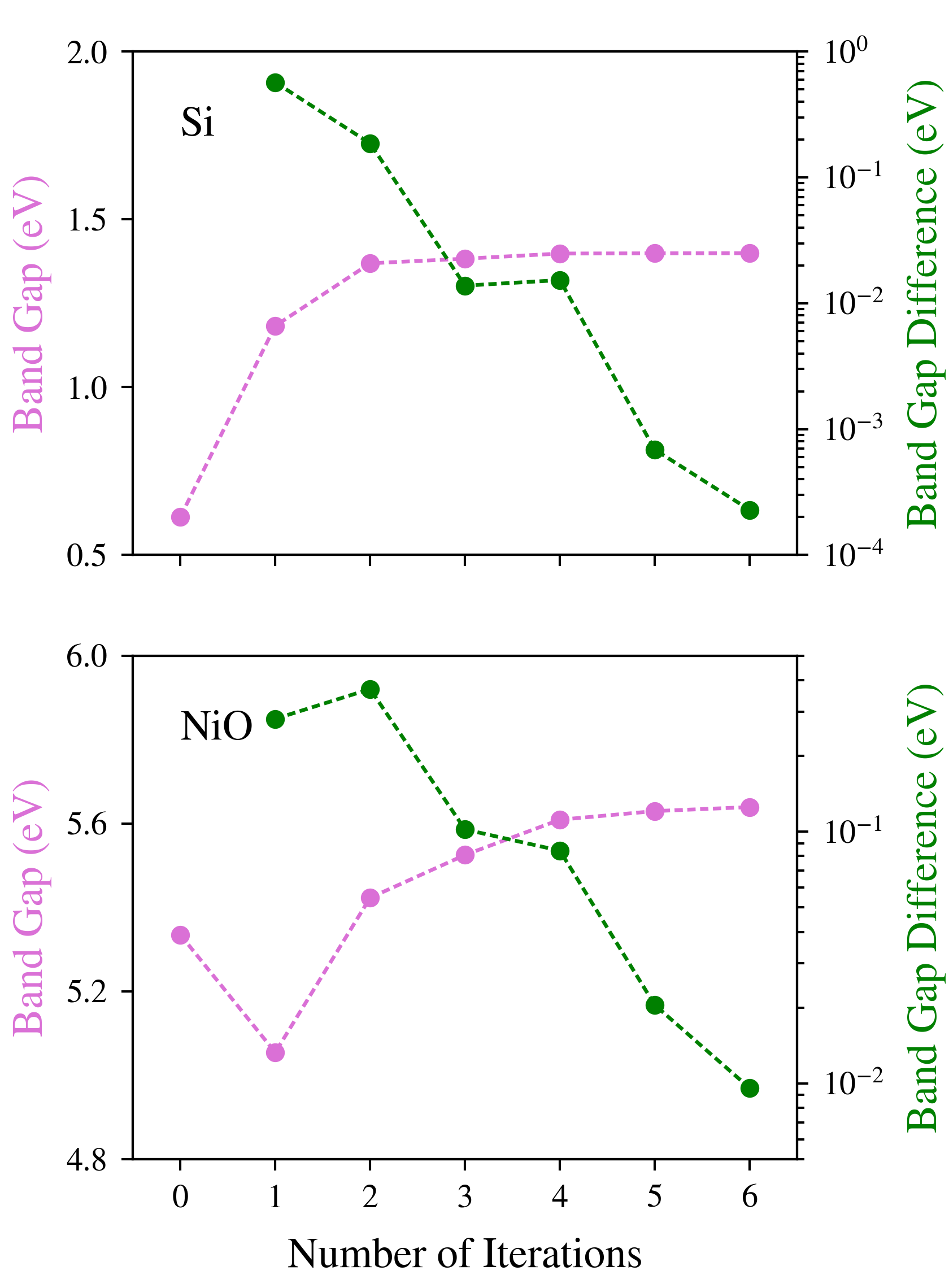}
\caption{Convergence of QSGW band gaps on Si (top) and NiO (bottom) with respect to the self-consistent iterations.}
\label{fig:converge}
\end{figure}

We first show the convergence of self-consistent QSGW calculations on Si and NiO in Fig.~\ref{fig:converge}. $6\times6\times6$ and $4\times4\times4$ $\veck$-point samplings were employed for Si and NiO respectively. A Coulomb divergence finite size correction scheme~\cite{Zhu2021a} was applied to the QSGW calculations.

For Si, we start the QSGW calculation from DFT molecular energies and orbitals computed with PBE functional~\cite{Perdew96PBE} (i.e., 0th iteration). As shown in Fig.~\ref{fig:converge}, the band gap of Si increases from the PBE value (0.61 eV) rapidly to the QSGW value (1.40 eV) within a few cycles. At 5th iteration, the band gap is already converged to within 0.001 eV. The QSGW convergence is slightly more challenging for NiO, even starting from a better DFT functional (PBE0)~\cite{Adamo1999}. Nevertheless, at 6th iteration, the band gap of NiO is converged to within 0.01 eV. For all other solids tested in this work, we find that 5-8 iterations are sufficient to obtain band gaps converged within 0.01 eV.

We also performed a numerical study on the starting-point dependence of QSGW using two DFT starting points: PBE and PBE0. As shown in Fig.~S1, for Si ($4\times4\times4$ $\veck$-mesh, no Coulomb divergence correction), both QSGW@PBE and QSGW@PBE0 predict the band gap to be 1.09 eV within 5 iterations. For more challenging NiO ($4\times4\times4$ $\veck$-mesh, no Coulomb divergence correction), both QSGW@PBE and QSGW@PBE0 band gaps are converged to 4.76 eV within 10 iterations. Thus, we have numerically confirmed that our Gaussian-based QSGW completely removes the starting-point dependence as expected.

Compared to one-shot \gw, the QSGW calculation is computationally more expensive due to two reasons: (1) multiple iterations are required to self-consistently converge QSGW calculations; (2) the full self-energy matrix needs to be computed in QSGW, while only the diagonal self-energy elements are needed in \gw. As an example, we find that our QSGW calculation on NiO ($4\times4\times4$ $\veck$-mesh) is roughly 19 times slower than the \gw@PBE calculation under the same numerical setting.

\onecolumngrid

\begin{figure}[hbt!]
\centering
\includegraphics[width=0.95\textwidth]{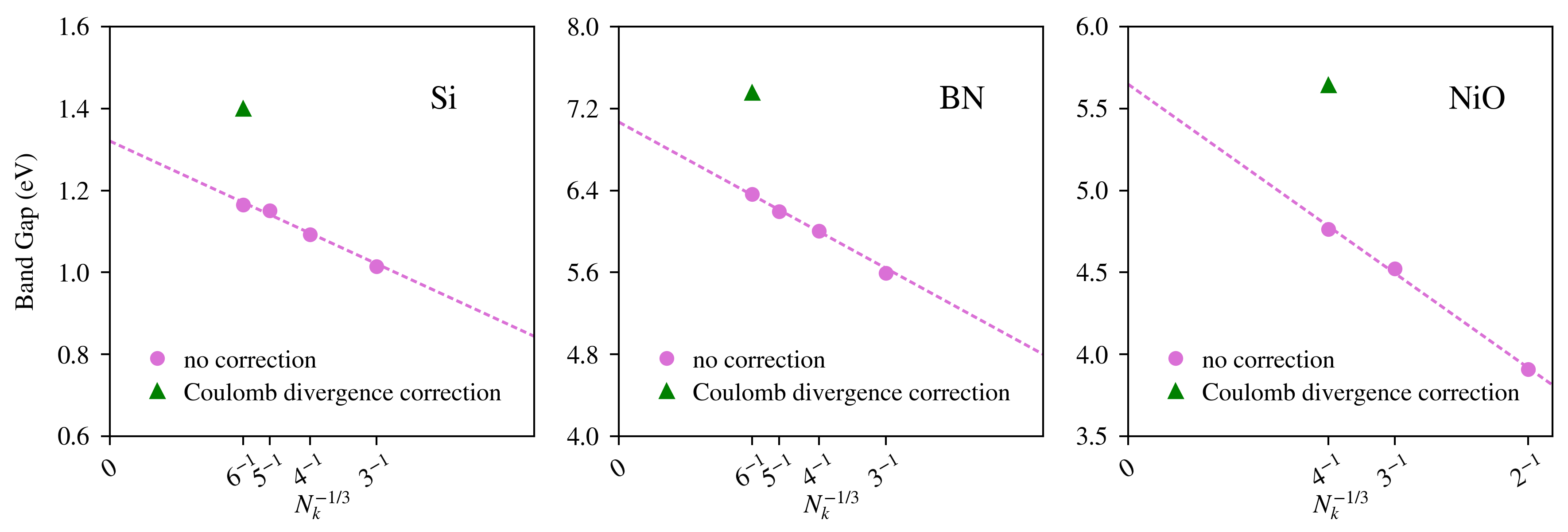}
\caption{Finite size extrapolation of QSGW band gaps with respect to $N_\veck^{-1/3}$ for Si, BN, and NiO. QSGW band gaps computed with the Coulomb divergence finite size correction are also shown.}
\label{fig:finitesize}
\end{figure}
\twocolumngrid

\subsection{Finite size correction}\label{sec:finitesize}
Next we test the finite size convergence of QSGW band gaps with respect to the size of $\veck$-point sampling on Si, BN, and NiO. For Si and BN, we calculated the QSGW band gap without finite size correction using $\veck$-meshes ranging from $4\times4\times4$ to $6\times6\times6$. We then conducted the finite size extrapolation to fit the band gap values according to the form:
\begin{equation}
    E(N_\veck^{-1/3}) = E_\infty + a N_\veck^{-1/3},
\end{equation}
to obtain the band gap at the thermodynamic limit ($E_\infty$). For NiO, $2\times2\times2$ to $4\times4\times4$ $\veck$-meshes were employed for a similar finite size extrapolation. As seen in Fig.~\ref{fig:finitesize}, consistently good fittings are obtained for QSGW band gaps of all three materials.

Using the finite size extrapolated band gap as a benchmark, we further study the accuracy of a Coulomb divergence finite size correction scheme developed in Ref.~\cite{Zhu2021a}, which was shown to perform well for Gaussian-based \gw@PBE calculations of solids. We find that, for Si, the Coulomb divergence corrected QSGW band gap computed at $6\times6\times6$ $\veck$-point sampling is 1.40 eV, which is 6\% overestimated compared to the extrapolated band gap (1.32 eV). Nevertheless, the Coulomb divergence correction still provides better accuracy compared to QSGW calculation without correction (1.17 eV at $6\times6\times6$ $\veck$-mesh). Similar behavior is observed for BN, where the Coulomb divergence corrected QSGW band gap at $6\times6\times6$ $\veck$-mesh (7.35 eV) is 4\% overestimated compared to the extrapolated value (7.07 eV). We speculate this slight overestimation in Coulomb divergence corrected band gaps is due to the incomplete basis set error in our correction scheme, which will be further studied in the future work. 

On the other hand, the Coulomb divergence corrected QSGW band gap of NiO at $4\times4\times4$ $\veck$-mesh is 5.64 eV, almost identical to the extrapolated band gap (5.65 eV), which we attribute to a better error cancellation in NiO with GTH-DZVP basis.

\subsection{Band gap and band structure of semiconductors}
We now present the band gaps of semiconductors and rare gas solids computed by Gaussian-based QSGW in Table~\ref{tab:bandgap}. For all cubic solids, we followed the finite size extrapolation procedure described in section~\ref{sec:finitesize}: we extrapolated QSGW band gaps (with no Coulomb divergence correction) computed at $4\times4\times4$ to $6\times6\times6$ $\veck$-meshes to obtain the thermodynamic limit value. We employed $3\times3\times2$, $4\times4\times3$, $6\times6\times4$ $\veck$-meshes for band gap extrapolation of wurtzite AlN (wAlN) and $3\times3\times2$, $4\times4\times3$, $5\times5\times3$ $\veck$-meshes for wurtzite ZnO (wZnO).

Compared to the experimental band gaps corrected for the zero-point renormalization effect, we find that Gaussian-based QSGW systematically overestimates the band gaps of semiconductors and rare gas solids. This is in constrast to the Gaussian-based \gw@PBE calculations performed with the same basis sets, where the band gaps are systematically underestimated. Both methods show dramatic improvement over DFT-PBE, which has a mean absolute relative error (MARE) of 43\%. Although QSGW is more computationally demanding than \gw, its MARE (10.1\%) is higher than \gw@PBE (8.2\%) for the tested materials.

We also compare our QSGW results against other QSGW implementations based on the linearized augmented plane-wave and muffin-tin orbitals (PMT)~\cite{Deguchi2016a}, projector augmented-wave (PAW) method~\cite{Grumet2018}, and the plane-wave (PW) basis~\cite{Chen2015} on semiconductors in Table~\ref{tab:bandgap}. Generally speaking, Gaussian-based QSGW results are in good agreement with these plane-wave based methods. The biggest discrepancy occurs in wZnO, which is known to be a particularly challenging material for converging \GW~calculations. We thus speculate this discrepancy results from the use of incomplete basis functions in some of the mentioned QSGW calculations. Nevertheless, we find our QSGW band gap for wZnO (4.63 eV) agrees well with the band gap of 4.61 eV in Ref.~\cite{Chen2015}. 

To quantify the basis set incompleteness error in our QSGW calculation of wZnO, we further conducted a numerical test using small $3\times 3\times 2$ $\veck$-mesh without Coulomb divergence correction. Three basis sets are tested: cc-pVDZ-pp, cc-pVTZ-pp, cc-pVQZ-pp for Zn, and cc-pVDZ, cc-pVTZ, cc-pVQZ for O. We find that the QSGW band gaps for wZnO are 3.44 eV (DZ), 3.35 eV (TZ), 3.41 eV (QZ). This result suggests, at least within the cc-pVXZ basis sets, the QSGW band gap of wZnO is converged to within 0.1 eV at the TZ level.

\onecolumngrid

\begin{table}[hbt!]
	\centering
	\caption{Band gaps of semiconductors and rare gas solids from DFT-PBE ($6\times6\times6$ and $6\times6\times4$ $\veck$-meshes for cubic and wurtzite structures), Gaussian-based \gw@PBE, Gaussian-based QSGW with finite size extrapolation, previous QSGW from literature, and experiments. Experimental values are corrected for the zero-point renormalization effect, as directly cited from Ref.~\cite{Ren2021} with the original values taken from Refs.~\cite{Madelung2004,Goldmann1989,Whited1973,Schwentner1975}. All band gaps are in eV. MARE stands for the mean absolute relative error.}
	\label{tab:bandgap}
	\begin{tabular}{>{\centering\arraybackslash}p{2cm}>{\centering\arraybackslash}p{2cm}>{\centering\arraybackslash}p{2cm}>{\centering\arraybackslash}p{2cm}>{\centering\arraybackslash}p{2cm}>{\centering\arraybackslash}p{2cm}>{\centering\arraybackslash}p{2cm}>{\centering\arraybackslash}p{2cm}}
	\hline\hline
	 &  & \gw@PBE  & QSGW  & \multicolumn{3}{c}{Previous QSGW} &  \\ \cline{5-7}
	System  & PBE  & (Ref.~\cite{Zhu2021a}) & (this work) & Ref.~\cite{Deguchi2016a} & Ref.~\cite{Grumet2018} & Ref.~\cite{Chen2015} & Expt.~\cite{Ren2021,Schwentner1975} \\
	\hline
    Si   &   0.61  &  1.08   &  1.32 & 1.28  & 1.49 & 1.47 & 1.23 \\
    C   &   4.14  &  5.52   &  6.14 & 6.11  & 6.43 & 6.40 & 5.85 \\
    SiC   &   1.36  &  2.44   &  2.81 & 2.63  & 2.88 & 2.90 & 2.57 \\
    BN   &   4.47 &  6.41   &  7.07 &   & 7.50 & 7.51 & 6.66 \\
    AlP   &   1.62  &  2.41   &  2.76 & 2.74  & 2.94 & 3.10 & 2.53 \\
    wAlN   &   4.19  &  5.89   &  7.09 & 6.91  &  & & 6.44 \\
    MgO\footnotemark[1]   &   4.75  &  7.43   &  9.33 & 8.97  & 9.58 & 9.29 & 7.98 \\
    wZnO   &   0.92  &  2.85   &  4.63 & 3.88  & 4.29\footnotemark[2] & 4.61 & 3.60 \\
    Ne   &   11.65  &  20.01   &  22.57 &   &  & & 21.7 \\
    Ar   &   8.77  &  13.24   &  14.88 &   &  & & 14.2 \\
    \hline
    MARE (\%) & 43.0 & 8.2 & 10.1 &  & & & \\
    \hline\hline
	\end{tabular}
	\footnotetext[1]{Most diffuse $p$ function of Mg was removed to avoid linear dependency.}
	\footnotetext[2]{Zinc blende structure.}
\end{table}
\twocolumngrid

\onecolumngrid

\begin{figure*}[htb]
    \centering
     \begin{subfigure}
         \centering
         \includegraphics[width=0.45\textwidth]{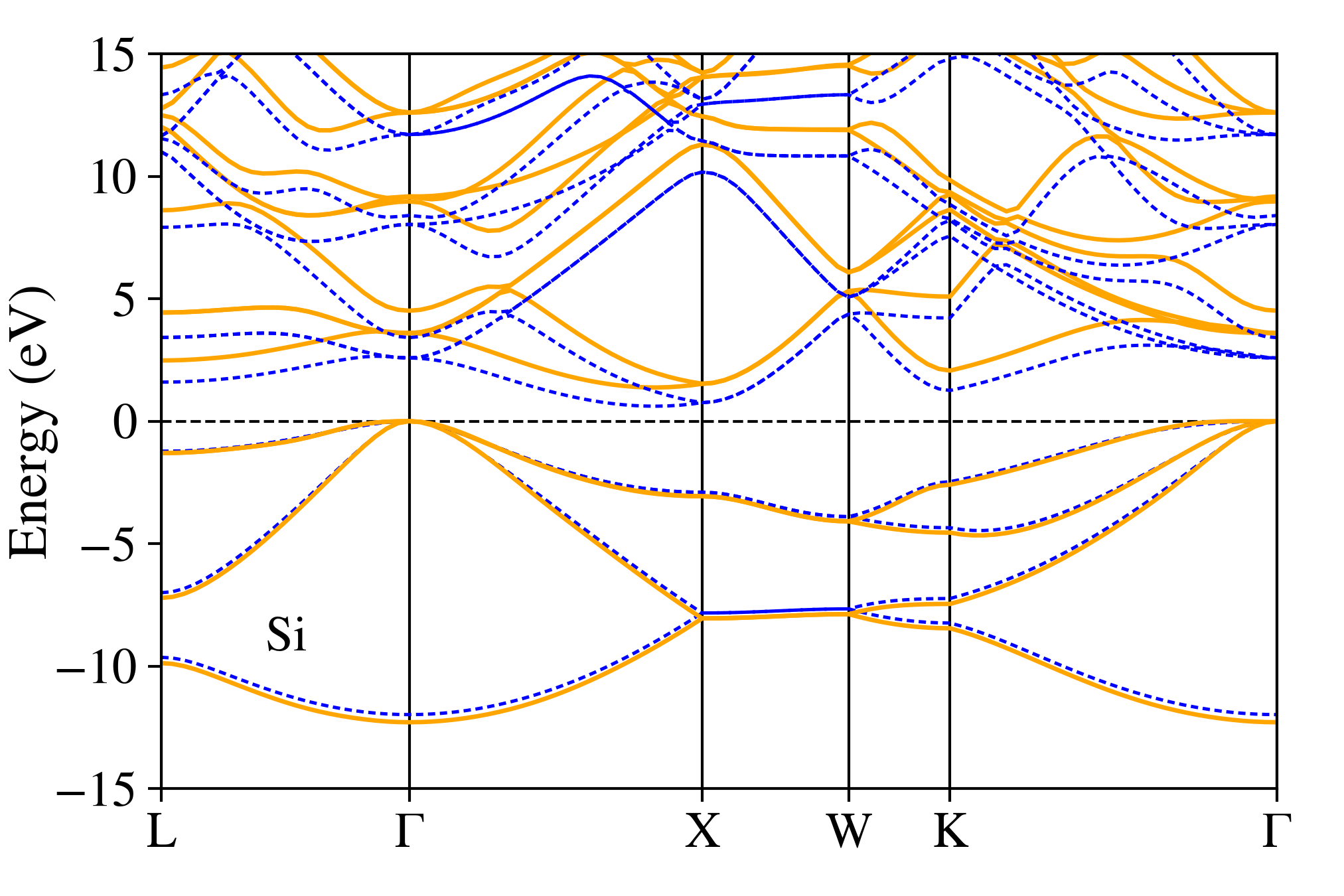}
     \end{subfigure}
     \hspace{1em}%
     \begin{subfigure}
         \centering
         \includegraphics[width=0.45\textwidth]{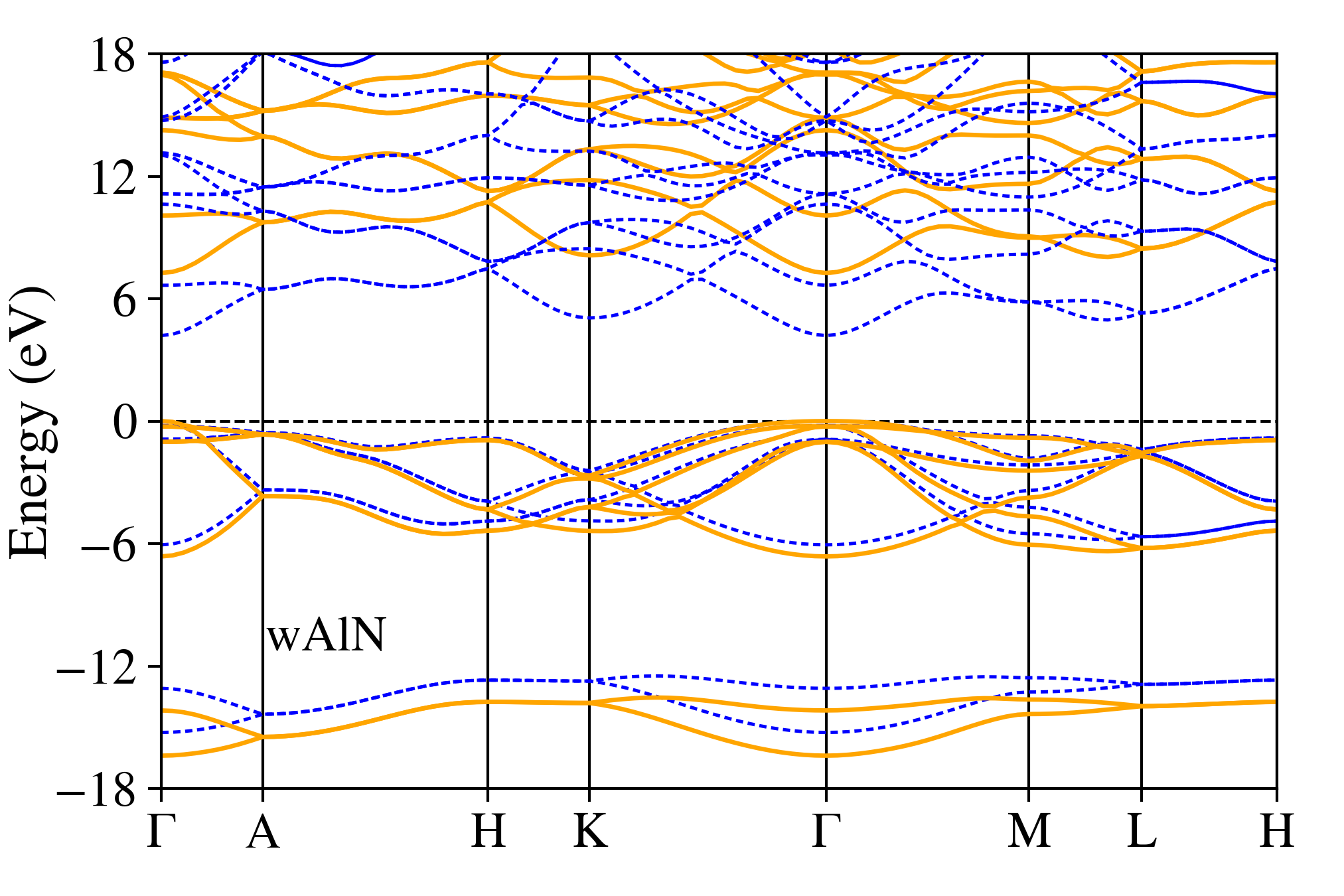}
     \end{subfigure}
    \caption{Band structure of Si (left) and wAlN (right) computed by QSGW (orange line) and DFT-PBE (blue dash).} 
    \label{fig:bandstructure}
\end{figure*}
\twocolumngrid

Furthermore, we calculated the QSGW band structure of Si and wAlN. $6\times6\times6$ and $6\times6\times4$ $\veck$-meshes were adopted for Si and wAlN, and Coulomb divergence finite size correction was applied. Following the Wannier interpolation method for \GW~QP energies~\cite{Hamann2009a}, we implemented a similar band interpolation scheme based on intrinsic atomic orbitals plus projected atomic orbitals (IAO+PAO)~\cite{Knizia2013d,Cui2020}. As can be seen in Fig.~\ref{fig:bandstructure}, QSGW consistently improves over DFT-PBE by predicting larger band gaps.

\subsection{Transition metal oxides}
Finally, we apply Gaussian-based QSGW method to investigate a series of prototypical strongly correlated materials: MnO, FeO, CoO, and NiO. To obtain band gap values in the thermodynamic limit, we performed finite size extrapolation using $\veck$-meshes ranging from $2\times2\times2$ to $4\times4\times4$.

\begin{table}[hbt!]
	\centering
	\caption{Band gaps (eV) of transition metal oxides in the antiferromagnetic phase.}
	\label{tab:tmo}
	\begin{tabular}{>{\centering\arraybackslash}p{1.1cm}>{\centering\arraybackslash}p{1.2cm}>{\centering\arraybackslash}p{1.2cm}>{\centering\arraybackslash}p{1.5cm}>{\centering\arraybackslash}p{1.5cm}>{\centering\arraybackslash}p{1.1cm}}
	\hline\hline
	 
    System &   PBE  &  PBE0   &  \gw@PBE & QSGW  & Expt. \\
    \hline
    MnO & 0.84 & 3.55 & 1.63 & 4.11 & 3.9$\pm$0.4~\cite{Elp1991} \\
    FeO & 0 & 3.15 & 0 & 3.67 & 2.4~\cite{Bowen1975} \\
    CoO & 0 & 4.27 &  0 & 4.83 & 2.5$\pm$0.3~\cite{Elp1991-2} \\
    NiO & 1.19 & 5.34 & 1.98 & 5.65 & 4.3~\cite{Sawatzky1984} \\

    \hline\hline
	\end{tabular}
\end{table}

As demonstrated in Table~\ref{tab:tmo}, we compare the QSGW band gaps with PBE, PBE0, \gw@PBE, and experimental values. Unlike for weakly-correlated semiconductors, \gw@PBE barely improves the description of strongly correlated TMOs over PBE. As a result, \gw@PBE is qualitatively wrong for the tested series of TMOs. For example, \gw@PBE predicts no band gap for FeO and CoO. In contrast, QSGW provides qualitatively correct prediction of band gaps for all TMOs, although the band gap values are overestimated. For MnO and NiO, the performance of QSGW is more satisfactory, where the band gaps are calculated to be 4.11 eV (MnO) and 5.65 eV (NiO). We note that these values are close to QSGW band gaps based on the PMT formalism (3.94 eV for MnO and 5.59 eV for NiO)~\cite{Deguchi2016a}. However, QSGW severely overestimates band gaps for FeO and CoO, suggesting low-order perturbation theory at the QSGW level is not enough to quantitatively describe FeO and CoO, which are known to be challenging correlated materials for \textit{ab initio} theories~\cite{Mandal2019}. We also find that PBE0 provides slightly better estimation of band gaps compared to QSGW on the tested TMOs.

\onecolumngrid

\begin{figure*}[htb]
    \centering
     \begin{subfigure}
         \centering
         \includegraphics[width=0.45\textwidth]{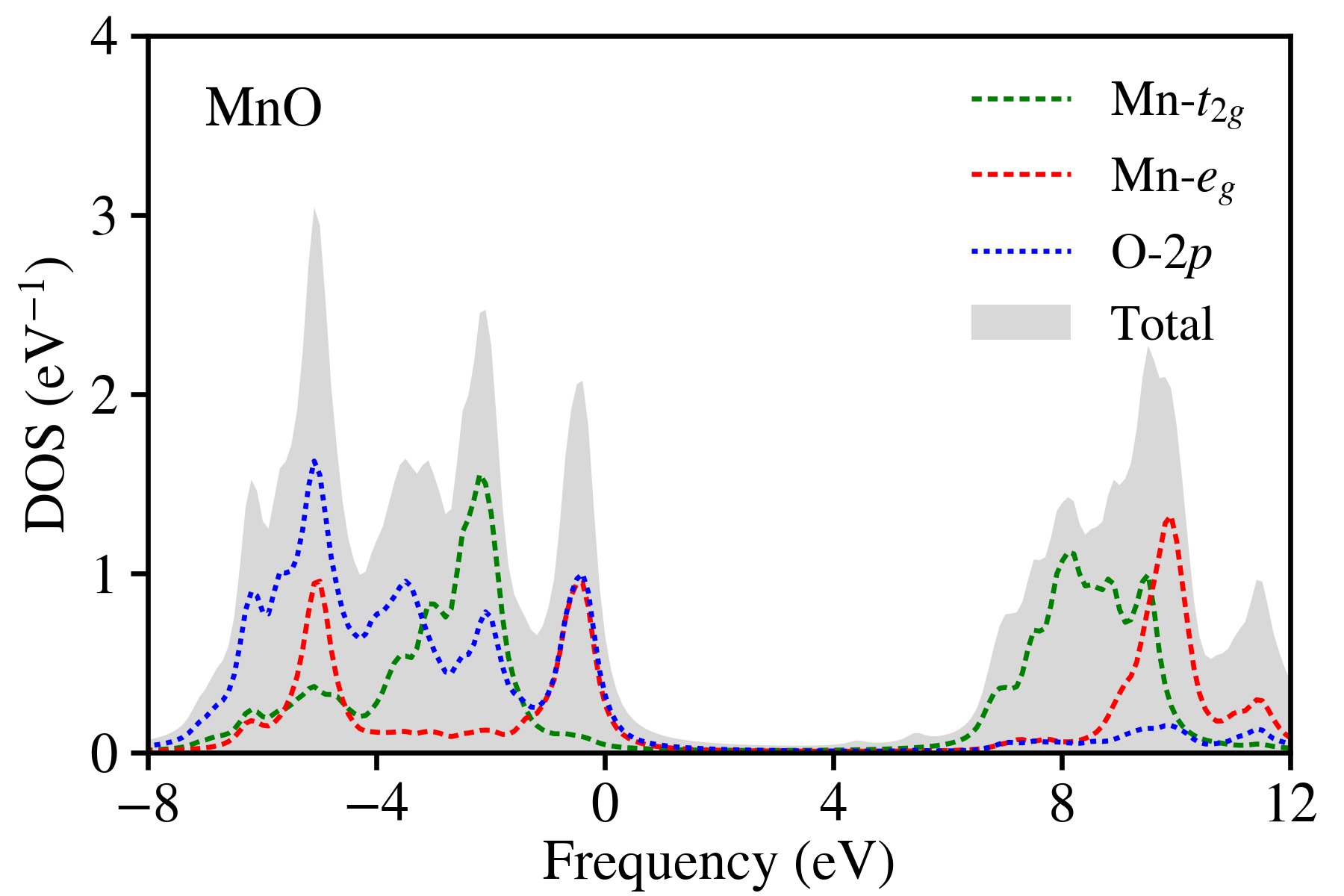}
     \end{subfigure}
     \hspace{1em}%
     \begin{subfigure}
         \centering
         \includegraphics[width=0.45\textwidth]{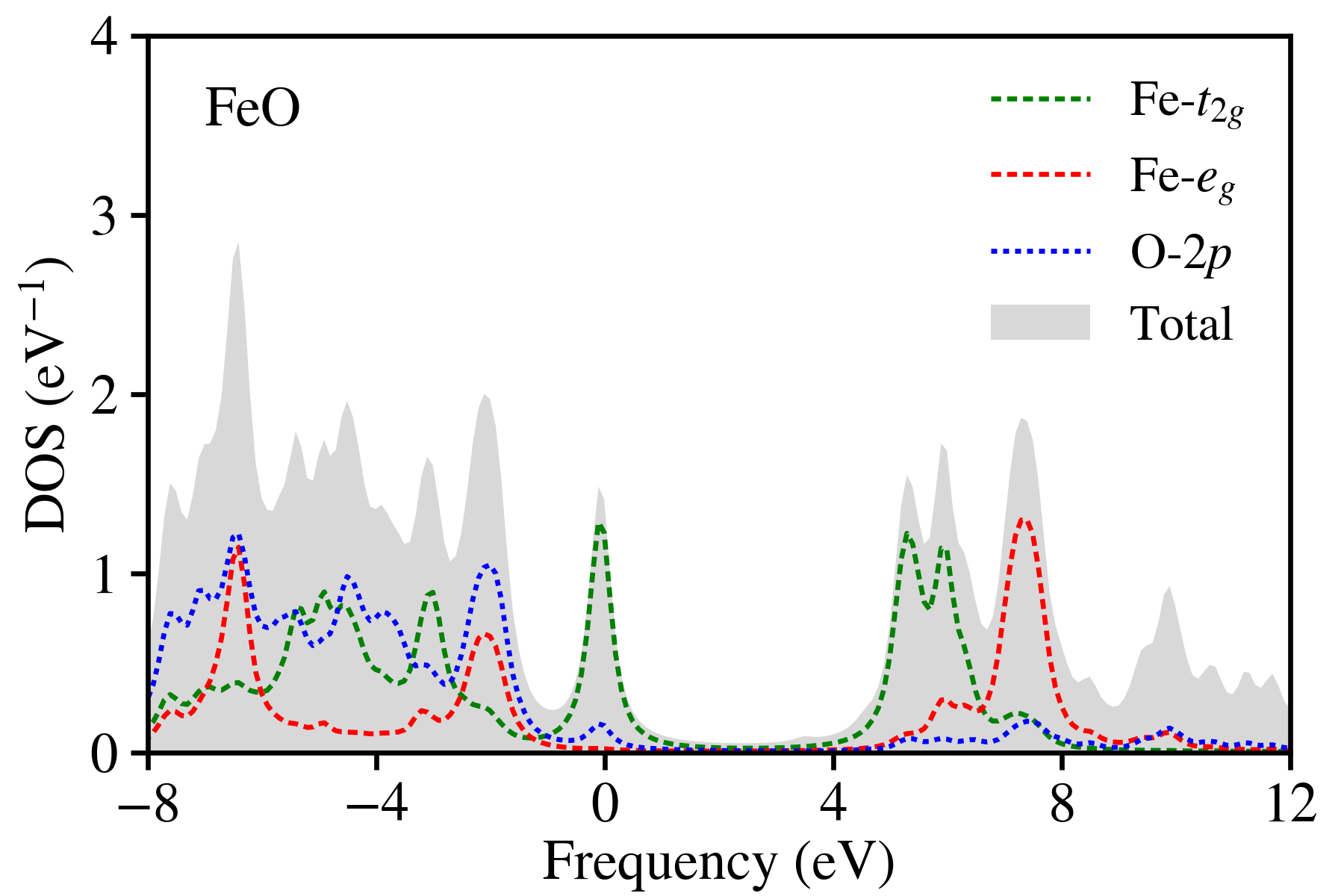}
     \end{subfigure}
     \vskip\baselineskip
     \begin{subfigure}
         \centering
         \includegraphics[width=0.45\textwidth]{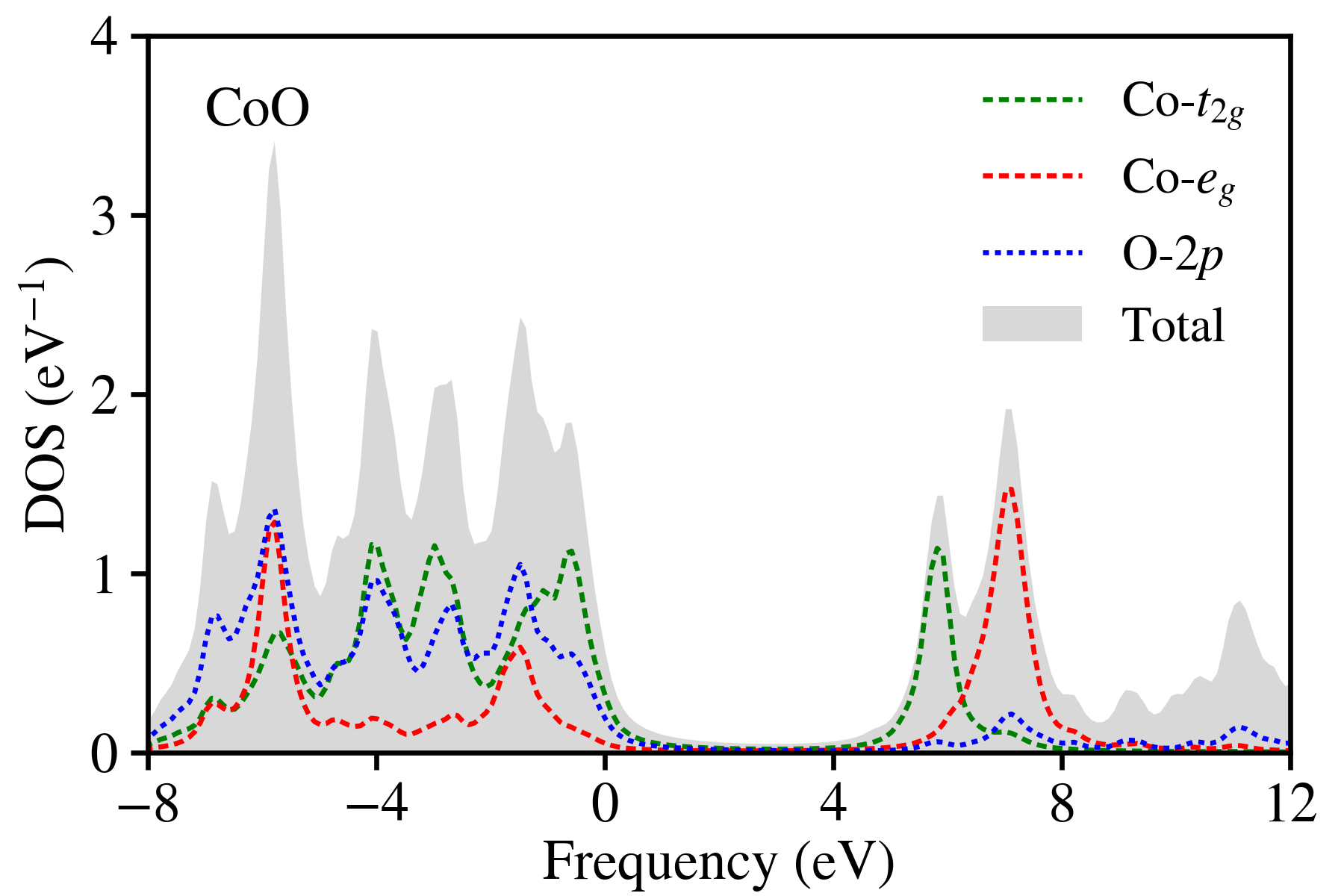}
     \end{subfigure}
     \hspace{1em}%
     \begin{subfigure}
         \centering
         \includegraphics[width=0.45\textwidth]{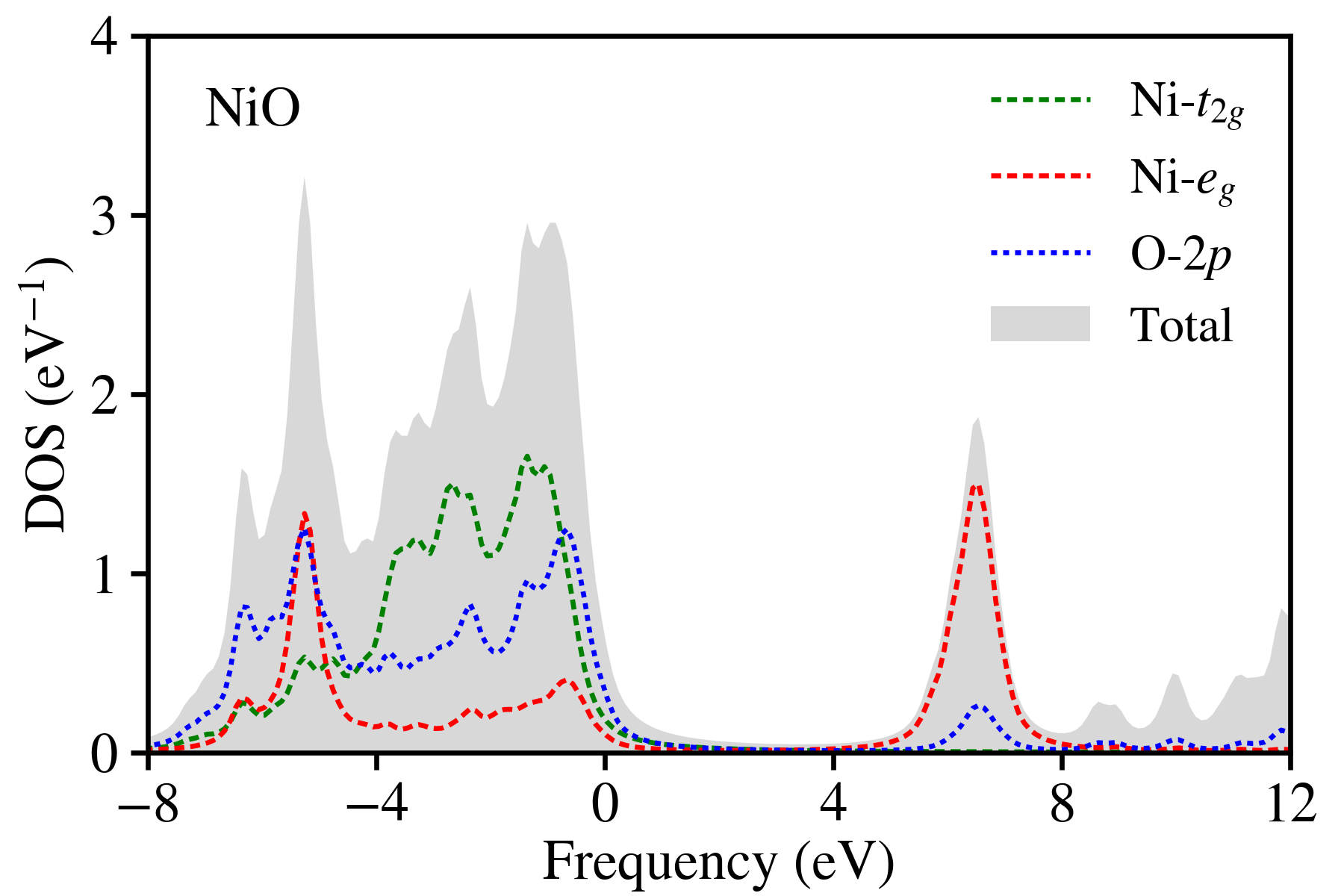}
     \end{subfigure}
    \caption{Orbital-resolved density of states computed by QSGW for MnO, FeO, CoO, and NiO.} 
    \label{fig:TMO}
\end{figure*}
\twocolumngrid

We further analyze the orbital-resolved density of states (DOS) of all four TMOs computed by QSGW, as presented in Fig.~\ref{fig:TMO}. Particularly, we find different insulating characters for the four TMOs. For MnO, the band gap is attributed to an excitation from Mn-$e_g$ and O-$2p$ to Mn-$t_{2g}$ orbitals. In FeO, while the lowest conduction band peak is of Fe-$t_{2g}$ character, the highest valence band peak is also mainly contributed from Fe-$t_{2g}$ orbitals. In both CoO and NiO, the highest valence band peaks have mixed contributions from metal-$t_{2g}$, metal-$e_g$, and O-$2p$. However, the lowest conduction band peak is of Co-$t_{2g}$ character in CoO but of Ni-$e_g$ character in NiO. 

When comparing the QSGW orbital-resolved DOS to our PBE0 calculations (Fig.~S2), we find very good agreement between two methods regarding the orbital components for all transition metal oxides. Similarly, our analysis on orbital components from QSGW agrees well with \gw@HSE03 calculations in Ref.~\cite{Rodl2009}, although there is quantitative difference in DOS peak positions and heights. Finally, we point out that QSGW qualitatively agrees with a recent \GW+DMFT calculation~\cite{Zhu2021c} on NiO that employs coupled-cluster Green's function as the impurity solver.

\section{Conclusions}
In this work, we present an implementation of periodic QSGW method based on crystalline Gaussian orbitals. We demonstrate that Gaussian basis set is a competitive choice for performing self-consistent \GW~calculations for periodic systems. Through benchmark on a series of semiconductors and transition metal oxides, we find that QSGW systematically overestimates band gaps for both weakly and strongly correlated systems, pointing to the need for further vertex corrections~\cite{Shishkin2007b,Kutepov2017a,Chen2015}. On the other hand, unlike the one-shot \gw~approach, QSGW completely removes the dependence on DFT functionals and provides more consistent descriptions over a wide range of solids. Overall, although QSGW is not a quantitatively accurate method for treating correlated materials, we believe it is promising to serve as a qualitatively correct, diagrammatically controllable starting point for high-level \textit{ab initio} quantum embedding methods~\cite{Zhu2019,Zhu2020,Cui2020,Zhu2021c}.

\begin{acknowledgments}
This work was supported by a start-up fund from Yale University. J.L. acknowledges support from the Rudolf Anderson Postdoc Fellowship. We thank Zhi-Hao Cui for helpful discussions and the Yale Center for Research Computing for supercomputing resources. This work also used the Extreme Science and Engineering Discovery Environment (XSEDE), which is supported by National Science Foundation grant number ACI-1548562.
\end{acknowledgments}

\section*{Author Declarations}

\subsection*{Conflict of Interest}
The authors have no conflicts to disclose.

\subsection*{Author Contributions}
\textbf{Jincheng Lei}: Conceptualization (equal); Data curation (lead); Formal analysis (equal); Methodology (equal); Software (supporting); Validation (lead); Visualization (lead); Writing - original draft (equal); Writing - review \& editing (equal). \textbf{Tianyu Zhu}: Conceptualization (equal); Formal analysis (equal); Funding
acquisition (lead); Methodology (equal); Software (lead); Supervision (lead); Writing - original draft (equal); Writing - review \& editing (equal).

\section*{Data Availability}
The data that support the findings of this study are available within the article.

%

\raggedbottom

\end{document}